**Design and synthesis of laser inscribed flexible Bragg gratings containing silicon oxynitride-doped silicon thin film for authentication applications with the insights from X-Ray Absorption Spectroscopy and Quantum Mechanical Calculations**

*Ali Karatutlu[1,2]*, Timuçin Emre Tabaru [2]*, Umut Taylan[1,4], Zehra Gizem Mutlay[1], Doğukan Hazar Özbey[1], Okan Uslu[2], Hamid-Reza Bahari[1], Esra Kendir Tekgül[1], Andriy Budnyk[1], İremnur Duru[2], Mustafa Fatih Genisel[4*], Zeynep Reyhan Öztürk[4,5], Taylan Gorkan[1], Engin Durgun[1*] and Bülend Ortaç[1]**

[1]Institute of Materials Science Nanotechnology and National Nanotechnology Research Center (UNAM) Bilkent University Ankara 06800, Türkiye

[2]Sivas University of Science and Technology, Mecnun Otyakmaz Street No:1 Sivas 58100, Türkiye

[3] Empa, Swiss Federal Laboratories for Materials Science & Technology, Laboratory for Advanced Materials Processing, Feuerwerkerstrasse 39, CH-3602 Thun, Switzerland

[4] Synchrotron-light for Experimental Science and Applications in the Middle East, Allan, Jordan

[5] Turkish Accelerator Radiation Laboratory (TARLA), Bahçelievler District 306 St. No: 11/N 06830 Gölbaşı, Ankara, Türkiye

Keywords: silicon oxynitride, flexible, waveguide, Bragg grating, femtosecond laser inscription, against counterfeit applications

E-mail: ali@unam.bilkent.edu.tr & akaratutlu@sivas.edu.tr, etabaru@sivas.edu.tr, mustafa.genisel@sesame.org.jo, durgun@unam.bilkent.edu.tr, ortac@unam.bilkent.edu.tr

Silicon-on-insulator is mostly main driving material system for the field of silicon photonics including PICs due to high index contrast and compatibility with current semiconductor manufacturing processes[1]. A recent example is silicon oxynitride-doped silicon[2] discovered recently by our group offering to set a new position for the fields of silicon photonics with a precise control of refractive index from 2.1 to 2.5 together with the photosensitivity property at different wavelengths. In this study, we demonstrate the synthesis of silicon oxynitride doped silicon can be performed on a flexible substrate for the fabrication of Bragg gratings (BG) at low temperatures below 100 °C that are emerging for advanced applications. Their use in preserving ultra-fast lasers inscribed micro-information has been demonstrated numerically and experimentally due to its multispectral reflection behaviour from the ultraviolet (UV) to the near-infrared (NIR) regions. The unique structural characteristics of silicon oxynitride-doped silicon were studied at the Si L, O K and N K edges using soft X-ray absorption studies. The quantum mechanical calculations provide the potential formation mechanism of SiON-doped Si whereas the crystal structure prediction for bulk $Si_3N_{4-x}O_x$ was conducted using the particle swarm optimization method in combination with density functional theory (DFT).

## 1. Introduction

A significant rise of phony products in electronics, pharmaceuticals, and luxury goods available on the market suggests that the verification of genuine origin remains an important and challenging issue for the industry. To embark upon this problem, state-of-the-art overt and covert anti-counterfeiting technologies are essential [5]. Nanotechnologies are advancing into the field by providing highly secure covert solutions based on purposeful variation of certain parameters at the nanoscale. For instance, epitaxial films of $VO_2$ on $Al_2O_3$ with Au electrodes were used as a binary encryption device controlled by multiple physical fields (voltage, temperature, and light) [6], the Ag plasmonic nanorods of varied length and orientation within the unit cell deposited on a Si wafer for polarization-controlled chrome-encryption [7], the ZnTe flakes photoluminescence ratio combined with electrically (the gate voltage) and optically (excitation laser) switchable logic gates encryption [8] and the encapsulated lanthanide doped inorganic matrix and layered hydroxide with PMA for the specific emission and color sequence multidimensional high-level information encryption [9]. The UV-excited optical authentication of ink stamps on cellulosic papers bearing photochromic and fluorescent latex nanoparticles, whose surface was doped with the spiropyran and coumarin derivatives [10].

Laser coding, nanoscribing and hologram creation have become important methods for adding security and anti-counterfeit measures. A femtosecond laser (fs-laser) allows precise processing of materials from ablation to the restructuring of the surface as it has been demonstrated for dental products (enamel, dentine, and cementum) [11], bulk metallic glasses [12], random metasurfaces [13], etc. The femtosecond-laser-induced phase transformations from the amorphous silicon (a-Si) to that crystalline (c-Si) and vice versa were occurring in thin films subjected to micropatterning, i.e. formation of either a 2D hologram image (a word ‘q-Psi’) or gratings by means of two-beam interference [14]. Inspired by this work, we recently reported on the low-temperature synthesis of amorphous silicon oxynitride-doped silicon (a-SiON-doped Si), whose alternating layers with silica ($SiO_2$) formed BGs on rigid substrates such as glass and silicon wafers [4]. BGs are reflecting structures with a periodic refractive index modulation. The proposed platform demonstrated high-reflection property at particular wavelengths with designed bandwidth values suitable for the authentication purposes. Such innovative strategy has been extended towards applications of photosensitive BGs in anti-counterfeiting techniques for wearable goods.

In the present work, the BGs includings a-SiON-doped Si were numerically designed and fabricated on a flexible substrates and exposed to the pulsed-laser light (fs and ns) for patterning. Because of the BGs' multispectral reflectance, the inscribed patterns (the letter "L" and the word "SPARK") can be observed at certain viewing angles or particular polarization of the incident light. Such unique feature opens an avenue for fabrication of the photosensitive BGs with the desired reflection wavelength that will guide the wearable item to be genuine. We demonstrate that an observer may check the authenticity of an item bearing a flexible BGs laser inscribed tag by simply varying the viewing angle (under indoor light) or by illuminating it with a conventional UV lamp (if no polarizer or lens are available).

## 2. Methods

### 2.1 Materials

Fused silica glass with a thickness of 1.1 mm and 300 nm $Al_2O_3$ coated 2" PVA films were used as the rigid and flexible substrates. $SiH_4$ (purity ≥4N) and $N_2$ (purity ≥4N) provided from a local company were utilized under nitrogen plasma in the plasma-enhanced chemical vapor deposition (PECVD) system.

### 2.2. Fabrication

Silicon-oxynitride doped silicon layer consisting of BGs on the PVA films were formed by the PE-CVD technique using a PECVD system (Model: Advanced Vacuum Systems) at relatively low temperatures, around 100 °C and a plasma power of 100 Watt (W) as described in our previous study [4]. The sample was immediately removed from the PE-CVD chamber and exposed to the ambient atmosphere.

### 2.3 fs-Laser inscription

For the fs-laser source, a Ti:Sapphire laser gain medium (Model: Spectra-Physics) was utilized operating at 800 nm ± 10 nm with a 150 fs pulse duration, 1 kHz repetition rate and 3 W of average power corresponding to 3 mJ of energy for each pulse. For the inscription of the flexible BGs, a couple of mirrors, a polarizer and a half wave plate were used to adjust the output power at 8 mW which was measured with a power meter before the fs-laser inscription process. The fs laser spot size was ca. 50 μm at the sample surface where the inscription was conducted. The shapes on the substrate were produced by a raster scanning process where the 8 mW laser beam was scanned with 5 mm length and 2 mm width and 2 mm length and 5 mm width rectangular shapes to produce the L shape. The scan speed was 1 mm/s and the pitch distance between lines of the raster scan was altered from 0.1 to 0.2 mm named with an increase of 0.05 mm after each inscription process. The respective inscribed points were named as P4 to P1 as shown from left to right in Fig. 3.e. This was important so that inscribed lines did not overlap or were not overly separated from each other. This also requires a careful adjustment of the lens height from the substrate and the laser power so as not to produce large depth of the inscription or stay below the inscription threshold. For the inscription of "SPARK" on the flexible BG, a commercial Q-switched ns laser was utilized with around 300 µm spot size and a power of 8.0 mW and 8.5 mW, respectively for AREA I and AREA II in Figure 7.

**2.4 Characterization**

SEM imaging was performed through the cross-section of the sample inside a focused ion beam (FIB) chamber of Nova NanoSEM 600i instrument. FIB was used to etch and then record the cross-sectional SEM micrographs of BGs.

Similarly XPS data on selected elements were acquired by a dedicated accessory after each etching cycle. During the XPS measurements to measure the depth profile of the flexible BG, an Ar ion beam was utilized with the etch rate of 0.3 nm/sec and total duration of 2000 sec corresponding to a thickness around 600 nm and 4 successive layers of SiON-doped Si and $SiO_2$.

For the TEM imaging, the laser inscribed flexible BG surface was exposed to a focused ion beam (FIB) within high-resolution scanning electron microscopy (HR-SEM, Model: FEI Nova NanoSEM 600i) to remove a thin film slice around 100 $nm^2$ surface area and over 1 µm depth. A platinum (Pt) protective layer was coated before this removal process. To mill perpendicular to the thin film surface using the dual beam of FIB, gallium ions were used at the energy of 30 keV. Then, the surface lamella was removed using a micromanipulator and transferred onto the copper (Cu)-coated carbon (C) TEM grid which was then taken for the TEM imaging.

A polariscope (Model: Optacore POL-02) was used to record the optical images of the BGs on the flexible substrates between two polarizer films under white light.

Optical UV-VIS-NIR spectra in reflectance mode were collected using a Carry 5000 UV-Vis-NIR Spectrometer ranging between 200 and 900 nm with a step of 1 nm and a one-beam configuration.

A PTFE reference provided with the instrument was used within the same measurement window for the baseline reflection.

The refractive indices of SiON-doped Si present in the flexible BG before and after the fs laser inscription were determined as described in our previous study[4] upon Kramers-Krönig model fitting[15] to the the dispersion relations Psi ($\psi(\lambda)$) and Delta ($\Delta(\lambda)$) measured by a variable angle ellipsometry (Model: J.A. Woollam Co. Inc.) at the angles of the incident light from 60° to 80° with 5° steps.

The FTIR measurements were performed at ATR (1500-400 $cm^{-1}$) region using an FTIR ATR spectrometer (Thermo Fisher Scientific). The Raman data was collected by a Raman spectrometer (Model: Witec SNOM Raman 300 Alpha) coupled with a diode laser at $\lambda$ = 532 nm using a grating with a groove density of 600 g $mm^{-1}$ for further structural analysis.  The optical images of the inscribed flexible BG were collected using the laser scanning optical microscopies equipped with a 100 W halogen lamp (Models: Keyence VKX 100 Microscope with 20x and 50x unpolarised objectives and Keyence VHX 1000 Microscope with a polarizing objective). A fiber coupled UV light source (Model: MIGREX FCS-0325-00) was utilized as a UV-LED light source at the wavelength of 340 nm. A 1 m long fiber (Model: Ocean Insight) with a core diameter of 600 microns to the light source. The center of the fiber were fixed with respect to the center of beam without the beam collimation so that whole sample were lightened by the sample. The angle-dependent reflection values were collected from the spots were the inscription took place using a 1 m long fiber m long fiber (Model: Ocean Insight) with a core diameter of 600 microns and the Ocean Optics-Flame mark spectrometer through the Oceanview software.

X-ray absorption (XAS) measurements were conducted at the HElmholtz-SESAME Beamline (HESEB) beamline in Synchrotron-light for Experimental Science and Applications in the Middle East (SESAME) in Allan, Jordan at the Si and N K edges at room temperature under vacuum environment below $1\times10^{-10}$ mbar. The SiON-doped Si samples (S1 and S2) were coated on a copper substrate to conduct the XAS measurements using the total electron yield (TEY) technique.

### 2.5. Numerical Simulations and Computational Methodology

A periodic Distributed Bragg Reflector (DBR) consisting of alternating high ($n_1 = 2.3$) and low ($n_2 = 1.45$) refractive index dielectric layers (5 pairs) were modeled using the 3-dimensional

Finite Difference Time Domain (FDTD) method (in Ansys Lumerical). Representing the laser-processed area, an asymmetric pattern (text geometry) altering the local refractive index contrast ($n_{1i} = 2.0$, $n_{2i} = 1.4$) was integrated throughout the entire depth of this multilayer thin-film structure. The simulation, defined by the PML boundary conditions, includes a reflection monitor (frequency-domain field and power monitor), a plane-wave source (λ=0.3μm to 2.5μm), a movie monitor, the DBR (structure group), and a transmittance monitor (frequency-domain field and power monitor), all positioned sequentially along the z-axis. To better analyze the interaction between light and the DBR, a mesh area consisting of smaller cells(0.01μm) has been defined around the DBR.

The computational calculations performed in this study were based on density functional theory (DFT) calculations, as implemented in the Vienna Ab initio Simulation Package (VASP) [16,17]. The potentials for Si, O, and N elements were represented using the projector augmented wave (PAW) method with an energy cutoff of 500 eV [18]. The exchange-correlation potential was approximated through generalized gradient approximation (GGA-PBE) [19]. To sample the Brillouin zone of the unit cell, a Γ-centered 6 x 6 x 8 uniform k-point mesh was generated using the Monkhorst-Pack scheme [20]. The atomic positions were optimized using the conjugate gradient method until the forces on each atom were less than 0.01 eV/Å. The convergence criterion for the ionic steps was set to $10^{-4}$ eV. The minimum energy path (MEP) approach was implemented for the determination of diffusion energy barriers.

The crystal structure prediction for bulk $Si_3N_{4-x}O_x$ was carried out using the particle swarm optimization method in combination with density functional theory (DFT), as implemented in the CALYPSO code [24]. The initial $Si_3N_4$ unit cell consisted of 12 silicon and 16 nitrogen atoms. In the structure prediction process, the number of Si atoms was fixed, while the oxygen concentration was varied with x = 4, 6, and 8 to model different compositions of $Si_3N_{4-x}O_x$. For each oxygen concentration, a population size of 20 structures was evolved over 20 cycles, resulting in a total of 400 candidate structures per composition.

## 3. Results and Discussion

### 3.1. Design, Fabrication and fs-Laser inscription of the BG device

A periodic BG consisting of alternating an asymmetric pattern (SPARK) altering the local refractive index contrast ($\boldsymbol{n_{1i} = 2.0}, \boldsymbol{n_{2i} = 1.4}$) was integrated throughout the entire depth of this multilayer thin-film structure representing the fs laser-processed area. Figure 1.A and B. Here, a light source covering the visible spectrum was sent through 5 pairs of laser-inscribed region, which was held perpendicular, and 45° to the reflection monitor. Figure 1.C and D demonstrates the results from the reflection monitor where the inscribed area is clearly visible when the sample is in the tilted condition (VIDEO S1). The numerical simulation of the BG optical properties after every cycle, and finished when the desired reflection spectrum has been reached [4]. Because of being formed by multiple layers of alternating materials with different refractive index, the obtained BG device is called a distributed Bragg reflector (DBR) with partially laser inscribed region, which has different reflection properties with respect to the as-prepared DBR region. The electric field distributions using the reflection monitor at two different observations conditions (perpendicular and tilted) are also shown in Figure S1, which are also consisted with the results shown in the video monitor.

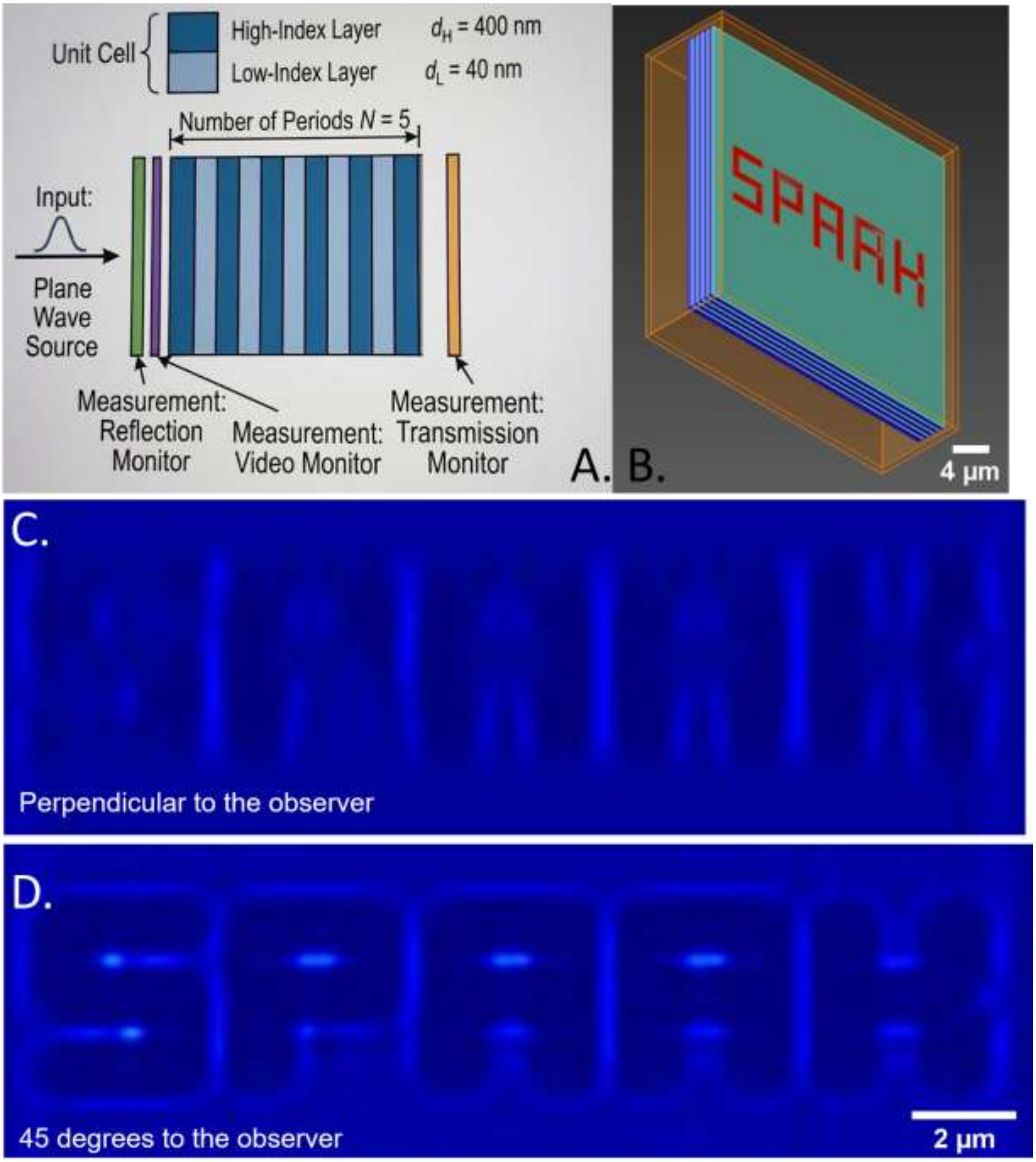

**Figure 1. A.** The design of the DBR and the locations of reflection, transmission and video monitors**, B.** the fs laser inscribed area "SPARK" (red in the turqoise) through the entire depth of the DBR**, C.** when the sample is perpendicular and **D.** 45° to the light source and the monitors.

For the proof-of-the-concept, the photosensitive BGs on a flexible substrate were designed as a self-twisting wearable ring. The amplitude and wavelength of its multispectral reflectance in the UV—NIR range were altered by the fs-laser inscription for the authentication purposes. Figure 2.a shows the schematic of the BG device fabricated in a plasma-enhanced chemical vapor deposition (PECVD) chamber at about 100 °C. The BG device consists of eleven repetitions of the SiON-doped Si photosensitive layers alternated by the $SiO_2$ layers grown atop of the alumina-coated PVA substrate.

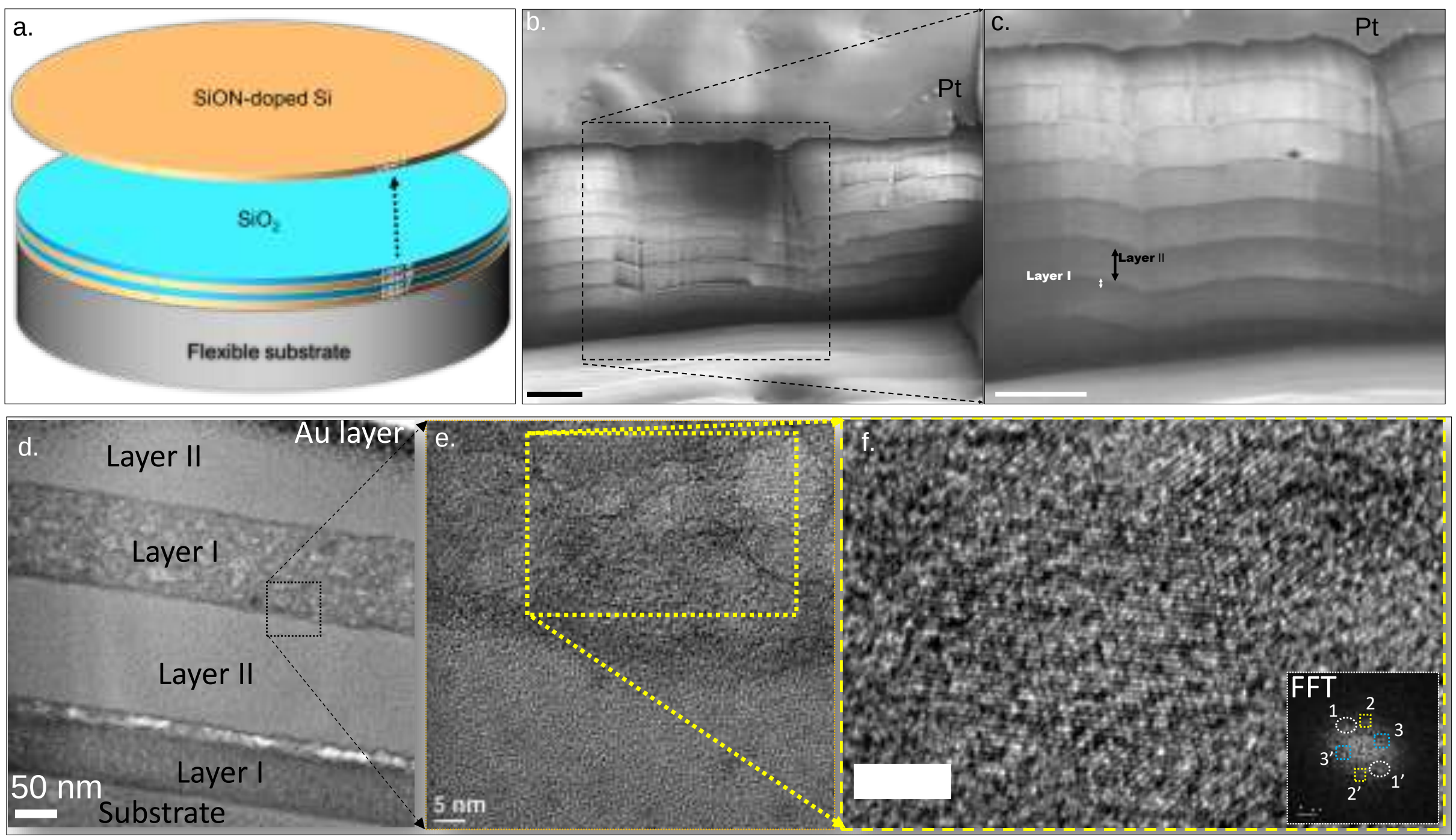

**Figure. 2. a.** Schematic of DBR made of 11 pairs of the SiON-doped a-Si and $SiO_2$ films on a flexible PVA substrate, **b,c.** The cross-section SEM images with magnification; the scale bars are 1 µm. **d.** The cross-section TEM image of the fs-laser inscribed DBR. **e.** The cross-section TEM image of SiON-doped a-Si layer, showing its partial crystallization under the fs-laser impact. **f.** The enlargement of **e** with the FFT image of the ordered area; the scale bar is 5 nm. The FFT image contains the indexed spots corresponding to the structural fringes within SiON-doped Si layer. The Au and Pt layers were coated respectively before and during FIB process to conduct the cross-sectional etching and cleaning processes.

The designed structure of DBR can be clearly observed at the cross-section of the as-prepared sample by scanning electron microscopy (SEM) as presented in Figure 2.b and c. Here, the morphological structure of the BGs deposited on a flexible PVA substrate was clearly observed with 11 subsequently deposited SiON-doped Si and $SiO_2$ layers. In the case of the PVA substrate, a wavy growth of the BGs was observed whereas the BGs were followed only by a horizontal deposition parallel to the top surface of the rigid substrate such as borosilicate glass whose structure was shown in Figure S2. This wavy growth appearing as a peculiar behavior could be due to structural deformation of the PVA substrate around 100 °C where the self-twisting is observed upon removing the PECVD chamber.

Due to the fabrication of the BGs at relatively low temperatures, the structure of BG becomes amorphous[4] and can be crystallized upon laser irradiation. An fs laser was utilized to test this phenomenon in the flexible BG. The cross-section of the laser crystallized flexible BG sample was partly removed by the focused ion beam etching and then further investigated using a transmission electron microscopy (TEM) imaging technique. Figure 2.d and e show that the cross-sectional image demonstrating the alternating thin film layers containing SiON-doped Si together with presence of the lattice fringes confirming the laser crystallization process. The Fourier transform on the image in (f) imaging provide information the corresponding selective area electron diffraction spots which will be correlated with the following in-depth structural analyses. Henceforth, the as-prepared BG on the flexible PVA substrate was called as the flexible BG for clarity**.**

### 3.2 In-depth structural analysis of DBR

Figure 3.a demonstrates grazing incidence X-ray diffraction (GI-XRD) of the flexible BGs before and after the fs laser inscription. The GI-XRD also confirms that the atomic structure of the as-prepared flexible BG is amorphous since a broad peak from 15° to 40° was observed. On the other hand, in the GI-XRD data, although some peaks seem to rise upon the fs laser inscription process, no clear assignment can be made due to the complex structure of SiON-doped Si since the broad peak still dominantly remains. SiON is indeed a broad term that encompasses a range of compositions between two extreme compounds, silicon nitride ($Si_3N_4$) and silicon dioxide ($SiO_2$), including various degrees of nitrogen and oxygen incorporation in an atomic level that indicates its intermediate nature. Nevertheless, after subtracting the broad baseline in the XRD pattern, the peaks can be assigned to the presence of $Si_2N_2O$ (Sinoite)[21,22], $\alpha$-$Si_3N_4$ [23] and c-Si[24]. The FFT data from the fringes of the TEM image (diffractional spots 1,1' and 2,2'-3,3' in the inset of Figure 1f) is provided in Table S1. The corresponding lattice spaces also confirms the existence of 111 planes of $Si_2N_2O$ and c-Si, respectively. Due to partial crystallization within the SiON composition range,

both amorphous and crystalline phases can exist. The structural features of the flexible BGs upon the laser irradiation appears to be preserved such as SiON-doped Si as demonstrated by their respective vibrational bonds in Figure 3.b. In our samples, either of amorphous or semicrystalline ones, we observe Si-O, Si-N, and Si-Si vibrations in FTIR and Raman spectra with the dominant atomic percentage of Si that denotes the existence of SiON solid solution basically within amorphous silicon. At the fingerprint region at around 760 $cm^{-1}$ and 902 $cm^{-1}$, Si-O-N and Si-N-Si bonds were respectively observed [25]. Figure 3.c shows the Raman data that compares as-prepared flexible BG, the laser inscribed BG and a reference c-Si in the region around 522 $cm^{-1}$ where symmetric stretching transverse optical (TO) phonon mode of c-Si appears [26]. Here, in the spectrum of the fs laser inscribed BG, together with a broad amorphous peak at around 475 $cm^{-1}$, a sharp peak at ca. 518 $cm^{-1}$ has risen whereas the as-prepared sample has only the broad amorphous peak located at around 475 $cm^{-1}$. The Raman data suggests that the fs laser inscription process on the flexible BG forms partially crystalline regions that introduce why the reflection alters under polarized light or at different viewing angles. This result is consistent with the previous study on the fs laser crystallization of pure a-Si [14]. Nevertheless, due to the unique reflection behavior of the fs-laser inscribed flexible BG fabricated in this study, the laser inscribed message can only be observed under certain conditions. Table 1 demonstrates the change in the refractive indices of the photosensitive SiON-doped Si before and after the laser inscription. The SiON-doped Si layer in the laser inscribed BG has relatively lower refractive index compared to the pristine SiON-doped Si layer. Therefore, the optical path length (OD) corresponding the geometric length (L) times the refractive index (n) of the medium ($\boldsymbol{OD}\colon \boldsymbol{L} \times \boldsymbol{n}$) has been altered due to the change in the refractive indices of the SiON layers in the laser inscribed BG. Since the geometric path appears to be preserved after the laser crystallization process, then the change in the refractive index of SiON-doped Si layers shown in Table 1 affirming to be responsible for the modification of the OD in the laser crystallized flexible BG with respect to the pristine flexible BG.

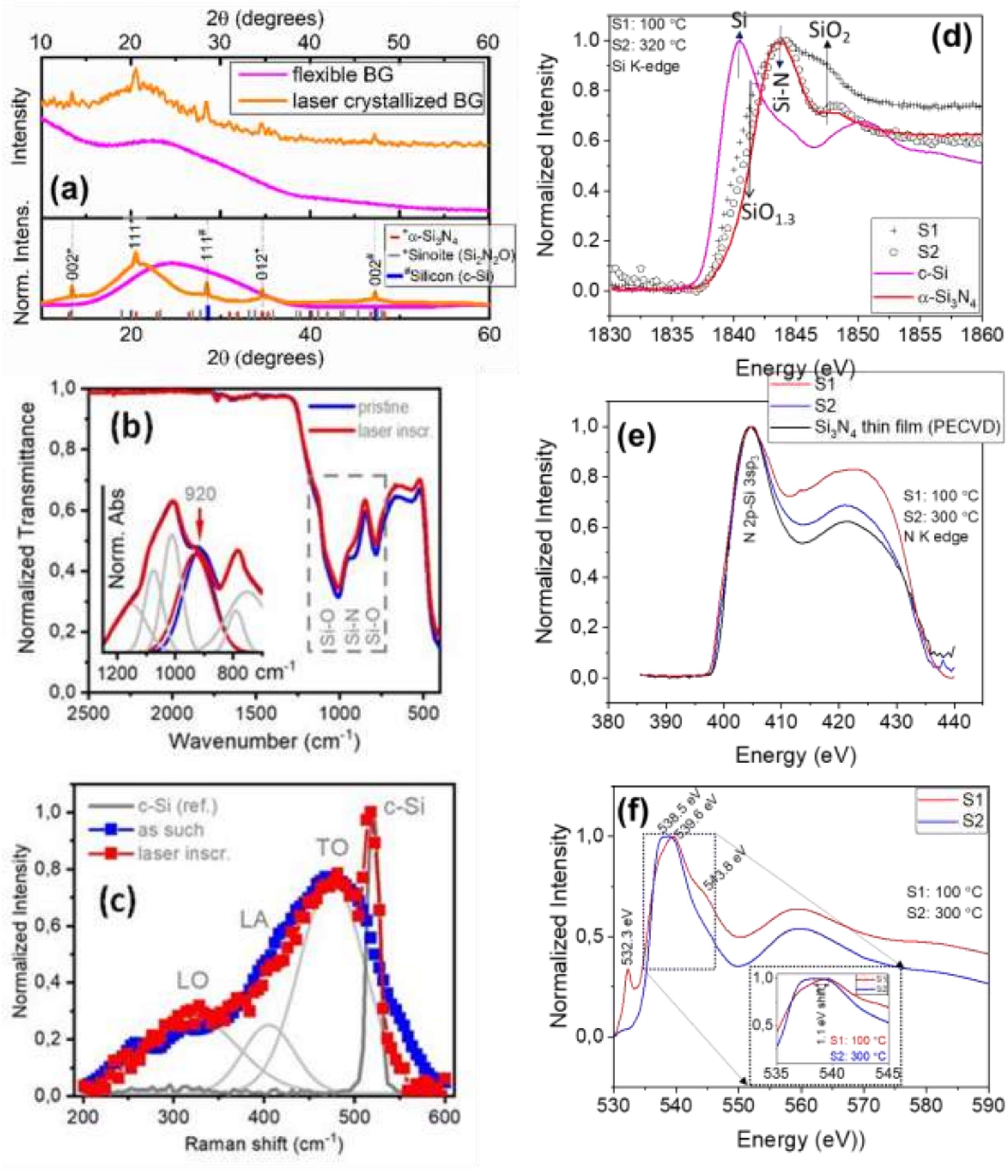

**Figure 3 a.** GI-XRD **b.** FTIR, and **c.** Raman spectra of the as-prepared and fs-laser inscribed flexible DBR. Additionally, the panel **a** reports baseline corrected XRD with reference data of sc-BG, the panel **b** demonstrates deconvolution of the Si-O-Si envelope, the panel **c** presents phonon components with a reference Raman data of c-Si. **d.** X-ray absorption spectrum of S1, S2, α-$Si_3N_4$ thin films and c-Si at Si K edge, **e.** at N K edge and **f.** at O K edge.

**Table 1.** Refractive indices of SiON-doped Si (@ λ=700 nm) before and after the fs laser inscription of the flexible BGs containing alternating layers of SiON-doped Si together with $SiO_2$. Here, the refractive index of $SiO_2$ is 1.38 @λ=700 nm.

| | N | K |
|---|---|---|
| SiO-doped Si as-prepared | 2.39 | 0.0792 |
| SiO-doped Si upon the fs laser inscription | 2.12 | 0.0006 |

Synchrotron radiation studies using X-ray absorption spectroscopy (XAS) technique was conducted via total electron yield (TEY) at Si K and N K edges to provide an additional view on the local structure of the SiON-doped Si within the flexible BGs. Fig. 3.d-e shows the normalized TEY intensity at Si K edge and N K edge for the SiON-doped Si synthesized at relatively lower temperatures around 100 °C similar to those within the flexible BG (S1) and at higher temperatures at 300 °C (S2). The TEY data in Fig. 3.d suggests that the structure of SiON-doped Si whose within the flexible BGs contains both stoichiometric and nonstoichiometric silicon oxides ($SiO_{1.3}$ and $SiO_2$) [27] in addition to the presence of Si and silicon nitride. On the other hand, N K edges shown in Fig. 2.e shows that the absorption edge does not appear to be altered. This suggest that silicon nitride within SiON-doped Si thin films does not appear to affected upon the temperature change despite the fact that the $Si_3N_4$ content is increased at lower temperatures (The XPS data in Table 1 in our previous study [4]). Nevertheless, Figure 2.f demonstrate a 1.1 eV shift in the O K edge of S1 with respect to that of S2 where nitrogen and oxygen contents are less. This infers that nitrogen-induced modifications in the electronic structure of S1. For example, relatively more SiON formation might disrupt O 2p – Si 3p orbital hybridization while reducing the localized empty states near conduction band edge causing to shift spectral weight higher energy transitions [28]. Based on our thorough structural analyses, Figure S3 can be proposed schematically to demonstrate that the as-prepared SiON-doped silicon within the flexible as-prepared BGs on the PVA substrate can include silicon, silicon dioxide, silicon nitride and silicon oxynitride structural units, separately. The structure of the SiON-doped Si using the lattice structure formed upon the laser inscription process and their possible formation mechanisms will be further be analyzed using the following studies of the quantum mechanical calculations.

### 3.3 Modelling of the SiON-doped Si structure within DBR

It is worth knowing that when SiON is in a crystalline (semicrystalline) phase, it often consists of a mixture of different crystallographic phases (mixed with amorphous) rather than a single, well-defined phase. This means that the crystalline phase of SiON can contain a combination of crystal structures from silicon nitride ($Si_3N_4$) and silicon dioxide ($SiO_2$), such as $Si_3N_4$ and $SiO_2$ crystallites coexisting within the same material. Since the formation of c-Si together with the amorphous SiON-doped a-Si upon the fs laser inscription process is evident by the Raman data, the formation mechanism can be considered to be consistent with the random mixing model (RMM) [29] where the model assumes Si, $SiO_2$ and $Si_3N_4$ can be randomly mixed.

To elucidate the potential formation mechanism of SiON within our system, we conducted quantum mechanical calculations based on ab initio computational methods. We started with the pristine crystalline phase (monoclinic) of bulk silicon nitride ($Si_3N_4$). The optimized lattice constants of the $Si_3N_4$ unit cell are found to be 7.81 Å, 7.81 Å, and 5.66 Å, which are in good agreement with the literature [30]. Subsequently, we examined the diffusion of a single oxygen atom within $Si_3N_4$ using a 2x2x2 supercell. Initially, it was found that oxygen (O) can settle in cage-like regions within the crystal without deforming the structure. Upon determining the adsorption site, one of the possible diffusion paths and its corresponding energy variation were calculated using the minimum energy path (MEP) approach, as depicted in Figure 4 (a-b). The highest energy barrier is identified as 0.37 for the initial step. Remarkably, once the oxygen initiates diffusion, the energy barrier decreases drastically and even drops below 0.1 eV. It is important to emphasize that the calculations were performed for the ideal $Si_3N_4$, and the barriers are anticipated to decrease in the structures with defects, particularly at elevated temperatures. Nevertheless, this analysis still indicates the plausible diffusion of O atoms into silicon nitride.

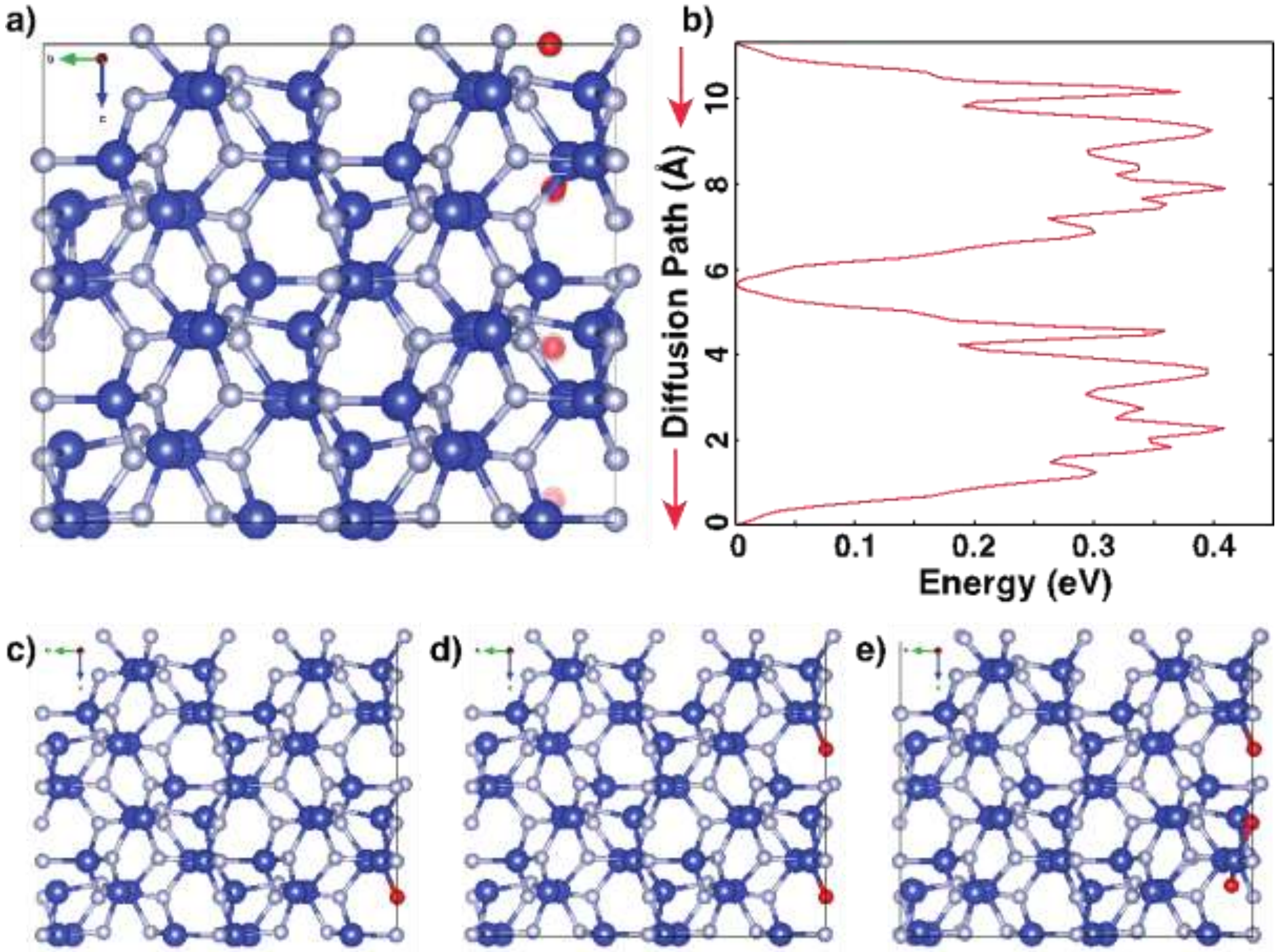

**Figure 4. a.** The illustration of the possible diffusion path of O within $Si_3N_4$ crystal and **b.** the corresponding energy variation. **c-e.** The representation of the O atom(s) substitution with N atoms in $Si_3N_4$ crystal where the number of O atoms were respectively increased from 1 to 3.

As a second step, we investigated the substitution of N with O atoms in $Si_3N_4$ and calculated the substitution energy ($E_s$). Es represents the energy required to replace N with O atoms. The average $E_s$ for addition of n O atoms can be calculated as:

$$E_s = \{E_T[\mathrm{SiON}] + \mathrm{n} * E_T\,[\mathrm{N}]) - (E_T\,[Si_3\mathrm{N}_4] + \mathrm{n} * E_T\,[\mathrm{O}])\}/\mathrm{n}$$

Where $E_T$ [SiON], $E_T$ [N], $E_T$ [$Si_3N_4$], and $E_T$ [O] correspond to the total energy of the O atom substituted $Si_3N_4$ crystal, single N atom, ideal $Si_3N_4$ crystal, single O atom, respectively. Within this definition, positive Es indicates that the substitution is an endothermic process. In this part, we considered n = 1 to 3 cases, as illustrated in Figure 4 (c-d-e). It is found that while Es for n=1 is calculated as 3. 29 eV, it decreases to 2.41 and 2.11 eV for n=2 and n=3, respectively. Different from the average, $E_s$ can also be calculated stepwise, where the substitution energy of $n^{th}$ oxygen is calculated with respect to (n-1). The energy required to substitute the second and third O atoms are found to be 1.54 and 1.49 eV, respectively. These results indicate that while a substantial energy is needed to substitute N with O atom, this energy notably decreases as the O content increases. It is important to note that the calculations take into account an ideal structure, and it can be more feasible for O atoms to occupy N vacancies followed by diffusion. Furthermore, the lowest $E_s$ values are obtained when the substituted O atoms are aligned along the same plane, resembling the O atoms in SiON. This alignment further signals a potential pathway for the formation of silicon oxynitride from silicon nitride in a medium with high oxygen content. Considering that the experimental conditions include the reaction temperature at around 100 °C and a plasma power of 100 W together with the excess amount of the oxygen source, such as $N_2O$ precursor, the synthesis of SiON-doped silicon requiring endothermic reaction can take place more likely.

To gain further insight into the formation of the SiNO crystal structure upon oxygen intercalation, we performed structure prediction calculations for $Si_3N_{4-x}O_x$ systems, using the parent $Si_3N_4$ structure as the starting point across varying oxygen concentrations (see *Computational Methodology*). The pristine $Si_3N_4$ crystal is composed of alternating Si and N atoms forming distorted hexagonal rings, as illustrated in Figure 5. Similarly, the predicted SiNO structure consists of dumbbell-like hexagonal rings.

Our structure prediction results suggest that oxygen atoms preferentially bond to nitrogen atoms located at the corners of these distorted hexagonal rings, thereby facilitating the formation of dumbbell-like configurations. The emergence of such ring structures is strongly dependent on oxygen concentration. At low oxygen content (e.g., $Si_3N_3O_1$), oxygen atoms tend to diffuse randomly throughout the lattice without establishing an ordered pattern. However, under oxygen-rich conditions, the possibility of forming ordered dumbbell-like SiNO hexagonal rings increases significantly. These dumbbell-like hexagons serve as the fundamental structural motifs of the SiON phase proposed in this study. Therefore, our findings indicate that the SiON crystal structure can emerge through the gradual bonding of oxygen atoms to the hexagonal ring framework inherent in the $Si_3N_4$ lattice, particularly under high oxygen chemical potential.

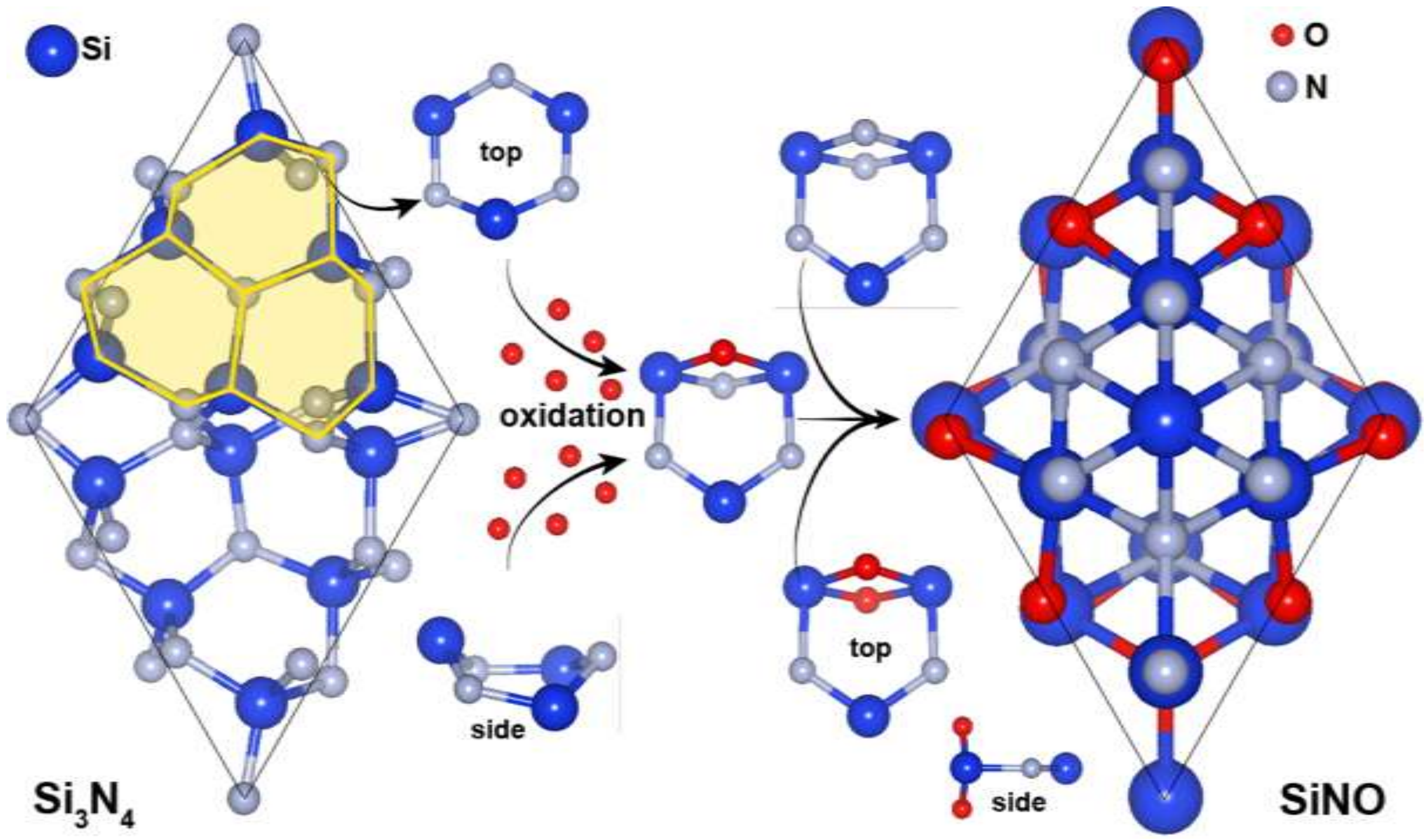

**Figure 5.** Schematic illustration of the SiNO crystal structure formation. Under oxygen-rich conditions, O atoms preferentially bond to nitrogen atoms located at the corners of the distorted hexagonal rings in the $Si_3N_4$ lattice. This bonding leads to the formation of dumbbell-like hexagonal rings, which serve as the fundamental structural units of the SiNO crystal structure.

### 3.4 Optical properties of DBR and authentication applications

The optical images of the BGs under different illumination sources including an in-door white light source, sunlight, and polarized light, are shown in Figure S2.a, b, and c, respectively. These light sources provide different features of the BG-coated PVA and demonstrate qualitatively homogeneous growth along the PVA substrate film. The polariscope image is also important to determine imperfections or patterns at different polarization angles that are not visible under normal lighting conditions. From the polariscope image, the flexible BG sample placed around a high-quality transparent glass rod is almost strain-free except where twisting causes a color change from brown to dark brown at the edges of the curvature.

Due to the multispectral reflectance behavior of the flexible BG, the message inscribed by the fs laser can be observed at specific polarization or different viewing angles, such as at 45° viewing angles under light exposure at different wavelengths (Figure 6 and Video S2). This includes the reflection image collected with a laser scanning optical microscope under an indoor white light with a polarized filter shown in Figure 6. a and b and under a UV light at different viewing positions revealed in Figure 6. d and e. Figure 6. c sheds light on the multispectral features of the as-prepared BGs under a broadband light source utilized by the diffused reflection spectroscopy (DRS) technique, and the redshift upon the fs-laser inscription demonstrates their versatility to serve as a signature of the inscription the letter “L”. In other words, the laser inscribed point has a relatively lower reflection compared to the as-prepared flexible BG with a redshift in all peaks in the spectrum. This feature can also be utilized as demonstrated in Figure 6. f, with the UV-LED source at three different viewing angles, including 45°, 60° and 90° with respect to the sample surface (0 °). This was also observed when keeping the observer and the light source constant but moving the flexible BG at different angles.

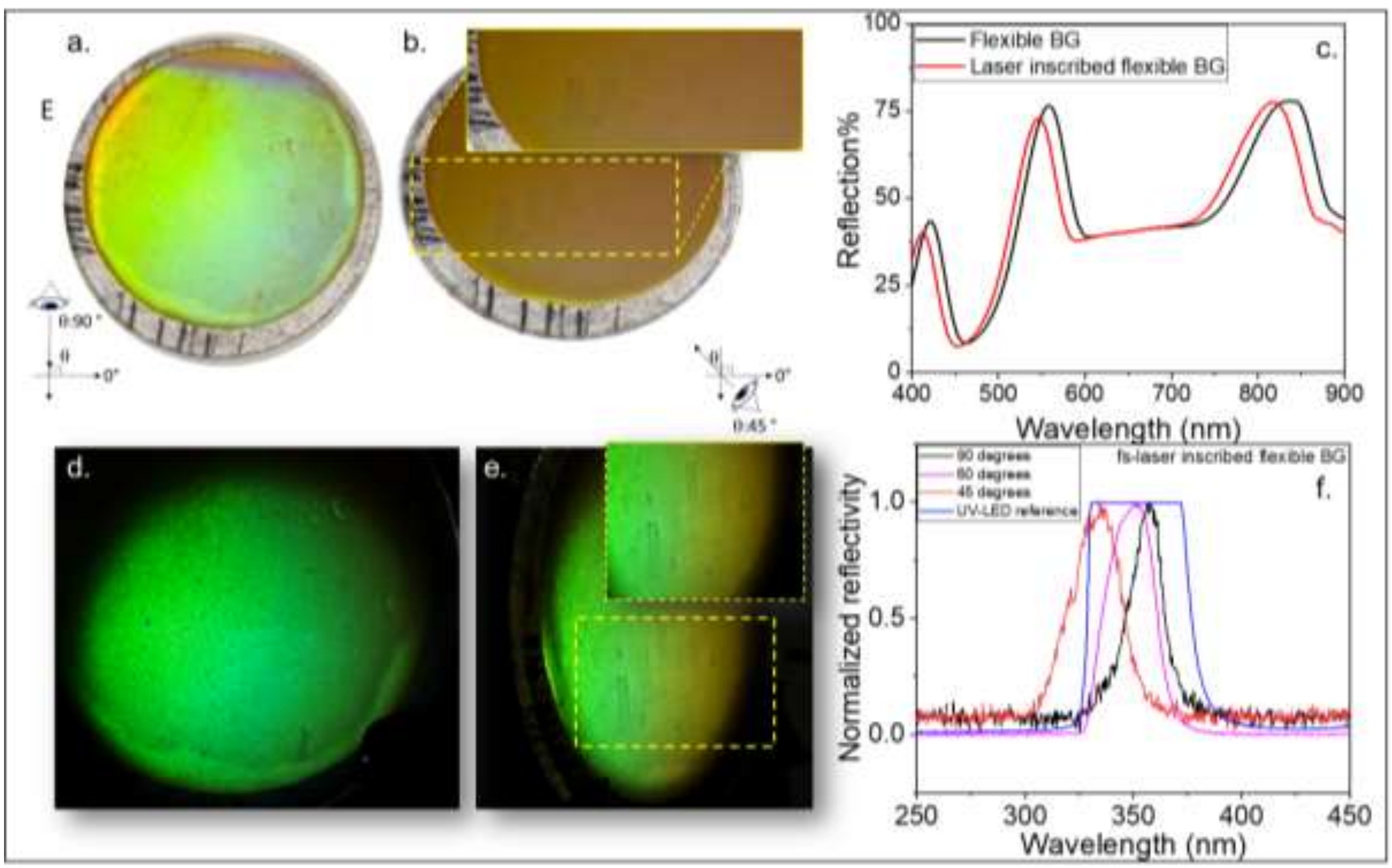

**Figure 6. a**. The optical images fs laser inscribed parts at different polarization under white light collected with a polarizing filter mounted onto a laser scanning optical microscopy and **c.** the reflectivity of the as-prepared and the fs-laser inscribed BGs collected when the samples were perpendicular to the light source and to the detector. Four L letters (P4 to P1, from left to right) are next to each other inscribed by different pitch distances from 0.1 mm to 0.2 mm. **d. e.**The optical images at two different observation angles (perpendicular and 45° viewing angles,

respectively) under UV-LED light recorded by iPhone 13 and f. the reflectivity data at the viewing angles of 45°, 60° and 90° compared with the spectrum of the UV-LED source.

Figure 7 shows this property revealed by a laser scanning microscope without the polarization filter scanned at the X, Y, and Z axes and moving the flexible BG perpendicular and 45° to the objective. The letter "L" appears vividly when the flexible BG is tilted at 45° (Figure 7. a, c, e) but starts disappearing when the sample is positioned parallel to the microscope stage (Figure 7. b, d, f). In other words, there is a slight difference in the reflection of the fs laser inscribed parts with unexposed parts. Therefore, when the sample was tilted the reflected power from the unexposed parts becomes relatively more compared to the laser inscribed parts which becomes darker and yield the observation of the inscribed message. The increase in the inscription length from 0.1 to 0.2 mm further reduces the reflection%, thus, it is important to choose an optimum value of the inscription length such as 0.1 mm to preserve over the desired reflection%.

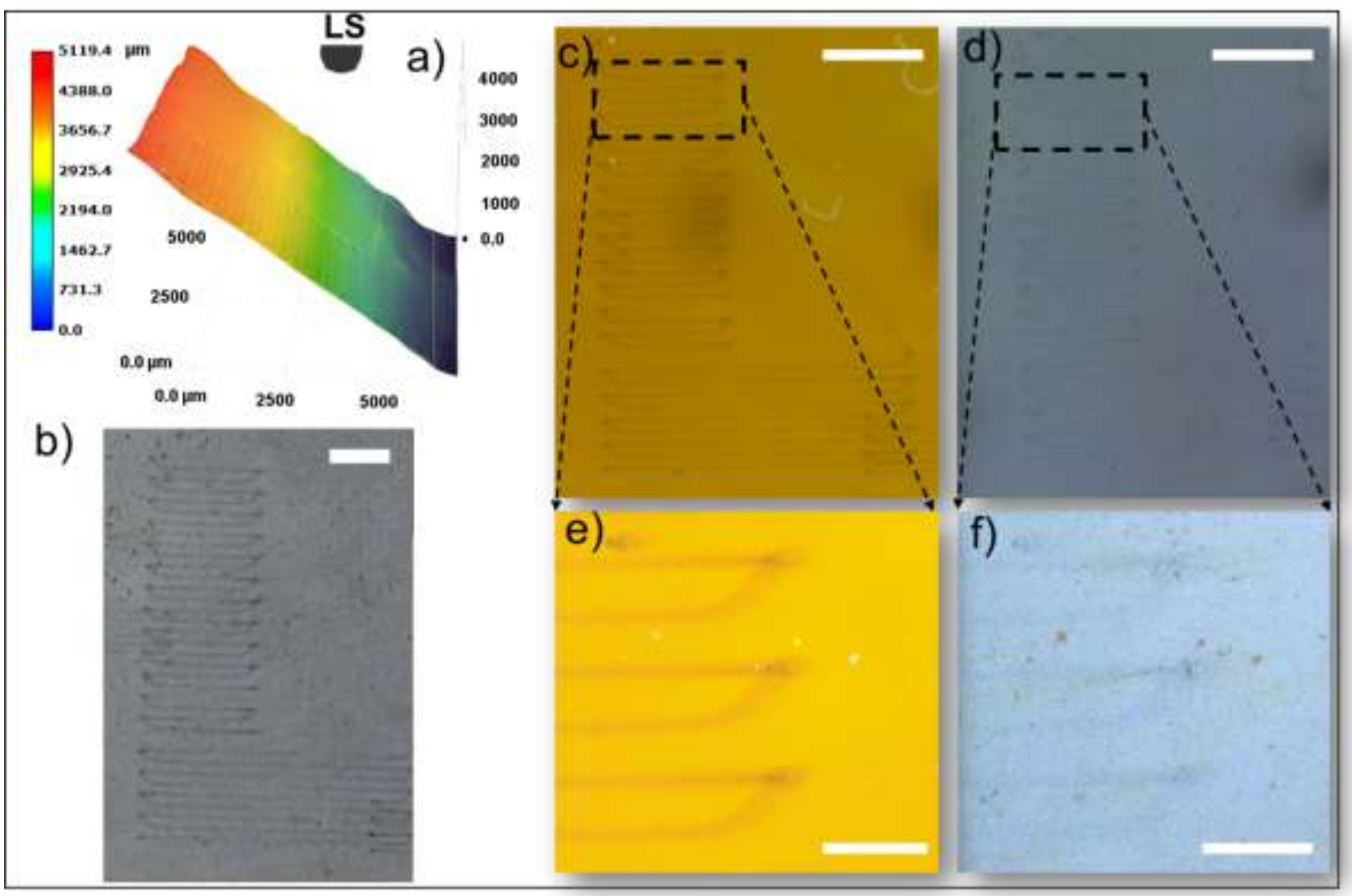

**Figure 7. a.** The optical images fs laser inscribed parts collected with laser scanning optical microscopy with relatively higher resolution at two different observation angles (**a, c, e** for 45°

viewing angle and **b, e, f** for 90° viewing angle) under white light using a halogen lamp source (LS). The scale in a demonstrates the difference in the height to show the 3D view of the surface at 45° viewing angle. The scale bars are 1000 µm in size. The corners of the inscribed letter "L" become more apparent due to the decreased inscription speed limited by the available stage in our laboratory. A 20x objective was used from a-d whereas 50x objective was utilized in e and f.

This feature can also be studies using an ns-pulsed laser. As an example, a newly designed flexible BG was fabricated on a wearable wool fabric having a periodic micropatterns (Figure 8). This time, a larger piece of the inscriptions "SPARK" were written using the ns laser on the BG at two different regions with a power levels of 8 mW and 8.5 mW. At relatively higher power levels, the fabric was initiated to be damaged. Figure 8 demonstrates the inscribed message "SPARK" can be observed at the polarization of 70 ° whereas at other polarization values, the letters are not clear and can become blurry. This demonstrates the inscribed BG structure at different flexible substrates consisting of the SiON-doped Si layers possess its unique behavior which is crucial to preserve the important information to be stored by the laser inscription that can be observed only at certain conditions including viewpoint of the observer or polarization of the incoming light.

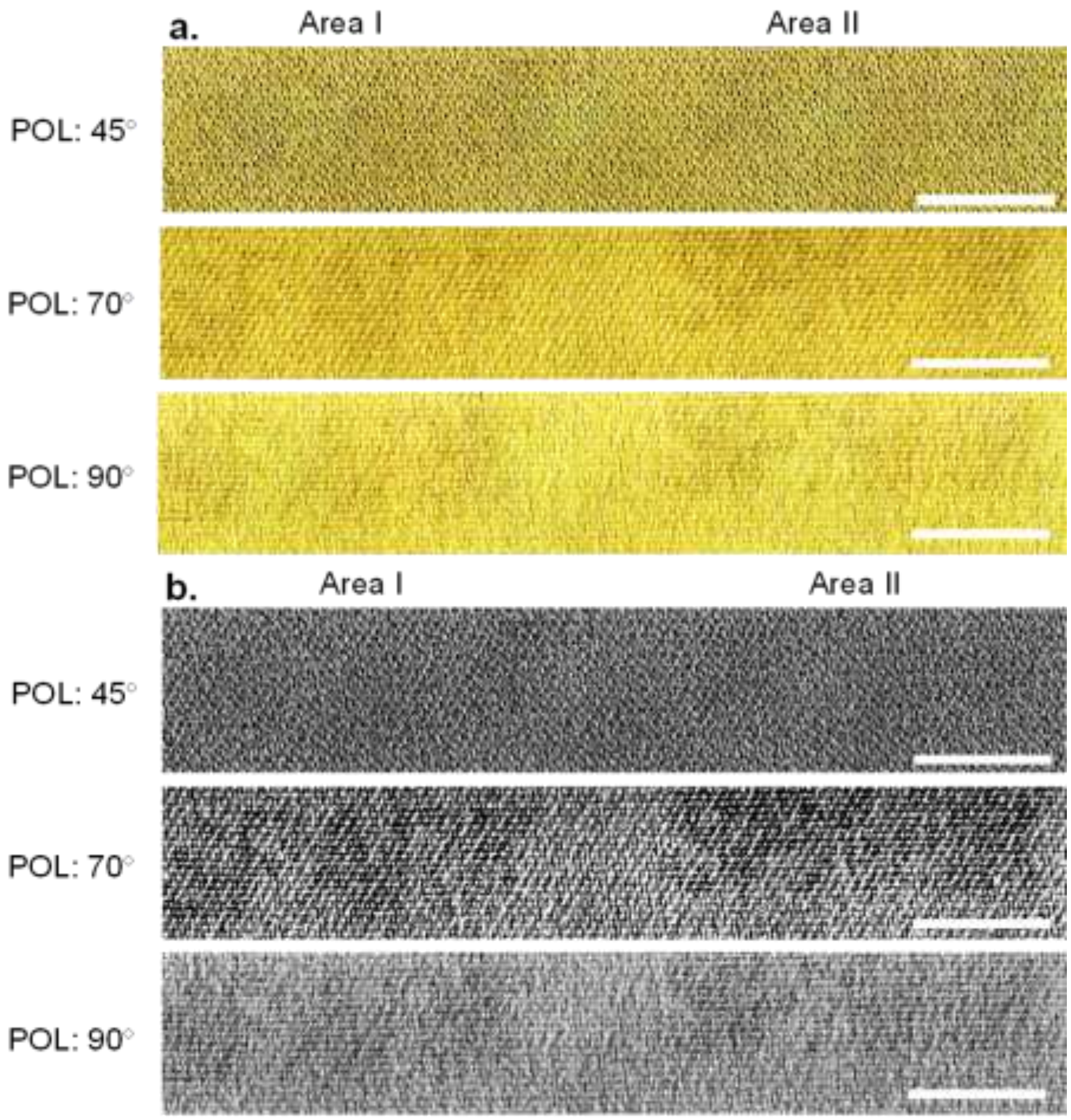

**Figure 8. a.** As-collected and **b.** the gray-scale optical microscopy images of the ns-laser inscribed message "SPARK" at two different areas and different polarizations (POLs ) including POLs: 45°, 70°, 90°. The scale bars are 5 mm in size.

## 4. Conclusion

Our study utilizing the fs laser inscribed flexible distributed BGs presents a novel but user friendly solution as unique host for any message, brand or images that are aimed to be protected. This innovative approach proposes high security via embedding the inscribed information within the nanoscale alternating SiON-doped Si and $SiO_2$ layers. On the other hand, it presents easily visible under diverse light conditions such as indoor and UV lamp environments. The inscribed message were written under the surface thorough the inner layers that further offers resistance to alteration and makes the flexible BGs reliable and effective in authentication applications. The synthesis of silicon oxynitride-doped a-Si via low-temperature PE-CVD technique was formed and shown for authentication applications upon the fs laser inscription on a flexible substrate. Due to the hyperspectral reflection property of the flexible BG, the laser inscribed can be monitored at different regions of the electromagnetic spectrum from the UV to NIR regions at certain viewing angles or particular polarization. The study is promising for preparing flexible BGs for wearable technology that can be protected with micro-information inscribed by a laser that can crystallize upon the irradiation and be observed only under the desired conditions, such as certain polarization of the incident light or the observation angles. The quantum mechanical calculations based on ab initio computational methods suggest that the SiON can be formed via the substation of the O atoms with the N atoms within $\alpha$-$Si_3N_4$ crystal and become more likely when the number of substituted oxygens increases. This possible bonding appears to cause to the formation of dumbbell-like hexagonal rings serving as the fundamental structural units of the SiNO crystal structure within the Si thin films present along the flexible BGs.

## Acknowledgements

The authors acknowledge infrastructural support provided by the National Nanotechnology Research Center. The authors acknowledge E. Y. Yıldırım for her help in collecting data when taking the optical images of the flexible BGs.

**Author contributions** These authors jointly supervised this work: Ali Karatutlu, Timuçin Emre Tabaru, Mustafa Fatih Genişel, Engin Durgun and Bülend Ortaç. A.K: conceptualization, methodology, investigation, data curation, funding acquisition, validation—original draft, review and editing. T.E.T: conceptualization, investigation, data curation (UV-Vis-NIR Reflection, XAS), funding acquisition, —, review, and editing. U.T: Investigation, data curation (fs laser inscription)— review, and editing. Z.G.M.: Investigation, data curation (Polariscope Imaging, XAS, SEM)—review, and editing. D.H.Ö: Investigation, data curation (DFT)— original draft, review, and editing. O.U.: Investigation, data curation (Numerical data)— original draft, review, and editing. H.B: Investigation, data curation (XRD)—review, and editing. E.K.T: Investigation, data curation (UV-LED reflection data)—, review, and editing. A.B: Investigation, data curation (XRD, FTIR, XAS)— original draft, review, and editing. A.B: Investigation, data curation (XAS)—, review, and editing. R.D.: funding acquisition, validation—review, and editing. M.F.G: conceptualization, investigation, data curation (XAS), funding acquisition, —, review, and editing. Z.R.Ö: conceptualization, investigation, data curation (XAS),—, review, and editing. D.H.Ö: Investigation, data curation (Crystal Structure Prediction)— original draft, review, and editing. E.D. and B.O.: visualization, investigation, funding acquisition, validation—review, and editing.

## Funding

This work was supported by the Scientific and Technological Research Council of Turkey (TÜBİTAK) ARDEB 3501 Career Development Program under Project No 121F255 and BİDEB-National Postdoctoral Research Program under the Project No. 118C541. Authors would like to thank Turkish Energy, Nuclear and Mineral Research Agency (TENMAK) for the travel grant to conduct the XAS measurements at the ID11L - HESEB station. The authors would like to thank the SESAME for the support of our proposal number of 20240171 to perform the XAS measurements.

## Competing Interest

The synthesis of silicon oxynitride-doped silicon, the waveguides, and the BGs prepared by this technique was the subject of a patent filing (Patent number: TR 2022/017824) (inventors: A.K., T.E.T., B.O.). The remaining authors declare no competing interests.

## Additional information Supplementary information

The online version contains supplementary material available at xxx.

**Correspondence and requests for materials** should be addressed to Ali Karatutlu, Timuçin Emre Tabaru, Mustafa Fatih Genişel, Engin Durgun and Bülend Ortaç.

Ali Karatutlu, ali@unam.bilkent.edu.tr , akaratutlu@sivas.edu.tr

Timuçin Emre Tabaru, etabaru@sivas.edu.tr

Mustafa Fatih Genişel, Mustafa.Genisel@sesame.org.jo

Engin Durgun, durgun@unam.bilkent.edu.tr

Bülend Ortaç, ortac@unam.bilkent.edu.tr